\documentclass[preprint,aps,showpacs,prc,nofootinbib]{revtex4-1}
\usepackage{graphicx}

\newcommand{\BV}{\left(\begin{array}{c}}
\newcommand{\EV}{\end{array}\right)}
\newcommand{\BM}{\left(\begin{array}{cc}}

\newcommand{\LQ}{``}
\newcommand{\bqn}{\begin{equation}}
\newcommand{\eqn}{\end{equation}}
\newcommand{\bqry}{\begin{eqnarray}}
\newcommand{\eqry}{\end{eqnarray}}

\begin{document}      

\title{ Entanglement and conserved quantities}
\author{T.\ Goldman}\email{tjgoldman@post.harvard.edu}
\affiliation{Dept. of Physics and Astronomy, \\
	University of New Mexico,\\
	Albuquerque, NM 87501 \\
{\rm and} \\
	Theoretical Division, MS-B283,
	Los Alamos National Laboratory, Los Alamos, NM 87545}

\begin{flushright}
\today \\
{LA-UR-22-22671}\\
{arXiv:2204.nnnnn}\\
\end{flushright}

\vspace{1.5in}

\begin{abstract} 
\begin{center}
The Einstein-Rosen-Podolsky paradox is resolved by reconsidering \\ 
what entangled state is actually prepared, what physical quantities are \\ 
conserved and the character of the correlation measurements employed. 
\end{center}
\end{abstract}

\maketitle

\section{Introduction}

The Einstein-Rosen-Podolsky paradox~\cite{EPR} focuses on the preparation 
and measurement of entangled states. The (almost universally referred to as 
EPR) paradox, as usually described, claims that if one of the two components 
of the entangled state is measured, then the other component, however far away, 
is then prepared by projection into a specific state even though a signal traveling 
from the measured component at the speed of light cannot reach it before a 
measurement of the other component is made. This would be a violation of special 
relativity. J.\ S.\ Bell~\cite{JSB,BA} sought to resolve this by proposing measurements 
of correlations, such as between light polarizations, that would reveal the existence 
of classical hidden variables that further characterize the entanglement. Since the first 
definitive experiment~\cite{CF}, efforts became more and more complex as theoretical 
arguments arose regarding subtle ways that the quantum information might have been 
transmitted despite the simple impossibility initially imagined.~\cite{Aspect, Zeilinger}  
The refinements in the experiments address all known conjectures as to how any 
communication might transpire between the two detectors displaying the quantum 
correlation of the entangled states, and all have definitively found values of the correlations 
that exceed those possible in classical mechanics (as described by Bell, with classical 
hidden variables) but which are in accurate agreement with quantum mechanics. The 
paradox thus remains as to how the \LQ later-observed" component has been projected 
onto the \LQ correct" state. 

I argue here that it is a combination of a conserved quantity, the state actually 
prepared and the nature of the measurement process that resolves the conundrum. 
{\it The quantum projections that do occur are entirely local.}

The first problem with the EPR view is that of timing. Consider the conventional case 
of two entangled photons created at the same instant, but departing the source in different 
directions. In the EPR analysis, either one of  the photons is measured first (in the rest 
frame of the source) while the other travels some distance away before being measured. 
However, if the \LQ second" photon is sent through a delay line before being released from 
the region of origin while the polarization of the first is measured, the process would 
be described simply as the preparation of the second photon in a prescribed state and 
there would be no surprise or dismay regarding the polarization measured for it later. 
Next, consider the simultaneous (again, in the rest frame of the source) measurement 
of both photons, where one cannot say that either one projected the other to the necessary 
correlated state or what direction any additional information (quantum or classical) travelled. 
These observations emphasize that something other or more than simply state 
projection/preparation is going on and that the understanding of the entangled state of 
the two photons used in these experiments is indeed deficient in some manner. An explicit 
example can clarify this point further. 

\section{Para-positronium}

Consider para-positronium annihilation as the source of the two photons. I choose this 
for simplicity, rather than a condensed matter source as, in principle, albeit with great 
experimental difficulty, one can imagine arranging for its natural decay by annihilation 
into two equal energy and oppositely directed photons to be accomplished very accurately 
in the rest frame of an optical system designed to carry off the photons to separate 
detectors. In a condensed matter source, one might need, in principle, to account for 
complications of phonons affecting the relation between the photons. In correlated 
sequential decays of an excited atom, there is also a (very small but not zero) time 
difference between the photon emissions, although neither of these concerns are viewed 
as significant in actual experiments.~\cite{{CF}, {Aspect},{Zeilinger}}

In this case, it is known that the initial entangled two photon state has total angular momentum 
zero, 
\bqn
J^2 = 0.
\eqn
My first observation is that the the entangled state actually completely prepared for observation 
of the two photons does not have this character after they enter the light \LQ guides".  By guides, 
I mean any of the following: fiber optics, a set of mirrors, lenses or simply the placement of the 
detectors that straight-forwardly collect events along a fixed $z$-axis. Lenses collect photons over 
a range of polar and azimuthal angles about the central $z$-axis, but momentum conservation 
requires each pair of photons to be back--to-back with the same character as those along the 
central $z$-axis. The alignment of the photons by the lenses from the different, independent 
$z$-axes to the central one simply collects them to arrive in parallel to the detectors at each end 
of the paths. In all cases, the $J^2$ of this particular entangled state has been redistributed over 
a wide range of values due to the collimation of the photons. Taking the common direction of motion 
to define the $z$-axis, the entangled state is projected into one of total $z$-component of angular 
momentum zero, 
\bqn
J^{tot}_z = 0 \; .
\eqn
This is a reasonable basis to choose as the photons are massless: They are naturally in chiral, 
or equivalently helicity, states. 

If we use this basis, the entangled state is described (where we take $p$ to be the modulus of 
the momenta along the $z$-axis, hence a positive quantity) as
\bqry
|\psi_e>  & = &  \frac{1}{\sqrt{2}}  ( |J_z=-1, p_z = +p>|J_z=+1, p_z = -p> \nonumber \\
          & + &  |J_z=+1, p_z = +p>|J_z=-1, p_z =-p>) \; ; 
\eqry
that is, the photons are either both left-chiral/negative helicity or both right-chiral/positive helicity. 
(Note the plus sign between the two terms as the interchanged objects are bosons.)  If we suppress 
the detailed momentum information, we can write this more simply as 
\bqn
|\psi_h> =  \frac{1}{\sqrt{2}} ( |L,-p >|L, +p> + |R, -p>|R, +p> ) \; .   \label{eq:chiral}
\eqn
(Again, the polar and azimuthal angles collected by lenses are irrelevant as linear momentum 
conservation requires that each detected pair is back-to-back along its own $z$-axis which each 
lens directs to its single common detector.)

It is immediately clear from this that, if both detectors have left-chiral pass filters or both have 
right-chiral pass filters, coincidences will occur at any detector separations/locations consistent 
with photon arrival times corresponding to the same entangled event. No state projection is 
involved as the effect of using two photon detectors with opposite chirality responses, such as 
described by $<L,-p|<R, +p|$, for example, is to annihilate the entangled amplitude, i.e., only 
one or the other photon could be detected, but there would be no coincident detections.  For 
that combination of detectors, both terms in Eq.(\ref{eq:chiral}) must (and do) include a factor 
of zero for the correlated detection amplitude of the \LQ other" photon. Otherwise, the detection 
of one left-chiral photon and one right-chiral photon would show that conservation of angular 
momentum had been violated. (And yet again, at whatever angle one lens collects a photon, 
the other lens can only collect the other photon at each of those angles plus $\pi$ due to 
momentum conservation.)

So whenever one detects the helicity state of one of the photons, it is angular momentum 
conservation that requires the helicity of the other photon to be the same -- no signal 
transmission is required in either direction as angular momentum conservation is always 
\LQ globally active", so to speak. Indeed, If $L$ were to be detected for one of the photons 
and $R$ for the other, one would have to conclude that some erroneous feature of the actual 
experimental setup introduced a violation of angular momentum conservation. On the other 
hand, it would be unusual, to say the least, to describe angular momentum as a hidden 
variable. Nor should this result be any more surprising than what would result  if the light 
guides (again, in the rest frame of the source) were not directed $180^o$ apart: If the opening 
angle were discernably more or less than that, one would certainly not detect (in the appropriate 
arrival time window) a photon in one of the detectors if a photon had been received in the other, 
due to conservation of linear momentum. 

In this example, there are no angular variations in correlations as discussed by Bell and 
examined in experiments, to test the inequalities he described; there is only an \LQ 
all-or-nothing" result.  Below, I examine those more general correlations and show that, in 
those cases also, conserved quantities are the origin of the observed quantum correlations 
which reflect the effect of a constraint, as above, i.e., a conservation law requirement that 
applies to each component of the full two particle entangled states initially envisioned in each 
case. This will reveal that the measured correlations are due to completely local measurements 
constrained by the conserved quantities.

\section{Massive particles with spin}

What is usually measured~\cite{Aspect} is transverse linear polarization, not chirality or 
helicity. We approach this in two steps, first considering transverse polarization of massive 
particles with spin. The two photon collision time reversal of the para-positronium decay 
slightly above the binding energy provides a conceptual reference although we consider 
instead, to avoid complications associated with electric charge or higher spins, a pair of 
identical spin-$1/2$ particles with no electric charge, but with a non-zero magnetic moment, 
produced entangled with zero total linear and angular momentum. 

For the sake of notational consistency, we continue to use $z$ as the axis defined by the 
momentum of the particles. The transverse polarization axis therefore lies in the $x$-$y$
 plane. If we consider the transverse polarization axis for each component of the wave function 
 at an angle $\theta$ corresponding to a pair of particles polarized at that angle to the 
 arbitrary $x$-axis, the value of $J^{tot}_{\theta}$ must be zero for each value of $\theta$. 
 The corresponding spin $+\frac{1}{2}$ (up) and spin $-\frac{1}{2}$ (down) non-relativistic 
 eigenspinors, $\phi$, in the $\theta$ direction are 
\bqn
\phi_{up\theta} = \frac{1}{\sqrt{2}} \left[
\begin{array}{c}  1 \\ e^{\imath \theta}  \end{array} \right]   
\eqn
and
\bqn
\phi_{dn\theta} =  \frac{1}{\sqrt{2}} \left[
\begin{array}{c}  -e^{-\imath \theta} \\ 1  \end{array} \right]  \;\; .
\eqn
These are the eigenstates of the spin operator $\sigma_{\theta} $, 
\bqn
\sigma_{\theta} = \frac{1}{2}\left[ {\rm cos}(\theta) \sigma_{x} 
                        + {\rm sin}(\theta) \sigma_{y} \right] 
\eqn
and they are the eigenspinors for each eigenvalue for all values of $\theta$ despite the 
unusual upper and lower component terms at values of $\theta$ far from $0$ or $2\pi$. 

The normalized entangled state is therefore
\bqry
|\psi_t> = \frac{1}{\sqrt{4\pi}}\int_{0}^{2\pi}  d\theta && \mkern-18mu \left[  
|\phi_{up\theta}, p_z = +p>|\phi_{dn\theta} ,  p_z =-p> \right. \nonumber \\
& - & \left. |\phi_{up\theta}, p_z = -p>|\phi_{dn\theta} ,  p_z =+p> \right]
\eqry
where, again, $p$ is a positive quantity. As we are  considering fermions, there is now a 
minus sign between the two terms at each value of $\theta$. The integral (sum over all 
orientations) is, of course, 
required by quantum mechanics due to the fact that the value of $\theta$ is not specified 
in the formation of the state, relative to any axis.

Since the direction of polarization is not specified by the entangled two particle emission state, we 
may choose an arbitrary transverse reference axis in the laboratory. As we envision the experiment 
to proceed by determining the polarization of each particle with a Stern-Gerlach apparatus, for 
example, we may choose the magnetic field axis of either Stern-Gerlach apparatus to define the 
$x$-axis. For each component of $\psi_{t}$, this will in general be at some angle to the polarization 
axis of the two particles defined by $\theta$. While the spin-direction detection must depend on this 
angle, the correlations between the detections of the two particles must obey angular momentum 
conservation. So if the spin polarization of one particle is at $\theta$ with respect to the polarization 
axis of one of the Stern-Gerlach detectors, the spin polarization of the other  particle must be at the 
angle $\theta \pm \pi$ with respect to that same Stern-Gerlach detector, that is, the latter must have 
the opposite spin eigenvalue. In this coordinate system, then, the entangled wave function therefore 
must have the form: 
\bqry
|\psi_t>  =  \frac{1}{\sqrt{4\pi}} \int_0^{2\pi} d\theta ( |J_{\theta}=+1/2, p_z = +p>
                          |J_{\theta} =-1/2, p_z = -p>  \nonumber  \\
                           -   |J_{\theta}= -1/2, p_z = +p> |J_{\theta}=+1/2, p_z = -p> ) \;  ,
\eqry
so that $J_{\theta}^{tot} = 0$ for every value of $\theta$. 

The correlation measurement proceeds by varying the magnetic field axis of the second 
Stern-Gerlach apparatus over angles $\omega$ with respect to that of the $x$-axis defined 
by the first. (Note that the polar and azimuthal angles collected by the lenses in the para-positronium 
case are analogous to $\theta$ here but have no effect since the spin polarization measured 
there is not transverse to the momenta of the photons.) 

\subsection{Correlation measurement}

We can now evaluate the correlated response of the two Stern-Gerlach detectors, the first  
with its magnetic field axis defining the arbitrary zero direction for $\theta$; we label this 
detector as $H0$. The magnetic field direction of the second detector will be varied through 
an angle $\omega$ with respect to the axis of the first to register the spin polarization 
correlation and is correspondingly labelled $H\omega$. The eigenspinors that are projected 
from the incoming particles are therefore, 
\bqry
\Phi_{H0up} & = & \frac{1}{\sqrt{2}} \left[ \begin{array}{c}  1 \\ 1  \end{array} \right]  \\
\Phi_{H0dn} & = & \frac{1}{\sqrt{2}} \left[ \begin{array}{c}  -1 \\ 1  \end{array} \right]   \\
\Phi_{H\omega up} & = & \frac{1}{\sqrt{2}} \left[ \begin{array}{c}  1 \\ e^{\imath \omega}  
\end{array} \right]   \\
\Phi_{H\omega dn} & = & \frac{1}{\sqrt{2}} \left[ \begin{array}{c}  -e^{-\imath \omega} \\ 1 
 \end{array} \right]   
\eqry
These spinors describe the particle that is detected at either the upper or lower exit of each 
Stern-Gerlach apparatus and so is registered as either spin-$up$ or spin-$down$ relative to 
the appropriate magnetic field axis. The amplitudes are therefore given by (where $<\Phi| = 
\Phi^{\dagger}$) 
\bqry
A_{updn\theta} & = & \{ <\Phi_{H0up}| \cdot <\Phi_{\omega dn}| \} \;  |\psi_t> \nonumber \\
       & = & \frac{1}{2\sqrt{2}} (1+ e^{\imath \omega} )  \label{eq:ud} \\
A_{dndn\theta} & = &  \{ <\Phi_{H0dn}| \cdot <\Phi_{\omega dn}| \} \; |\psi_t> \nonumber \\
       & = & \frac{1}{2\sqrt{2}} (-1+ e^{\imath \omega} )    \label{eq:dd}
 \eqry
Note that the results are independent of $\theta$. This depends crucially on the minus 
sign between the two terms in $\psi_t$ and that the $H0$ operation projects out a non-zero 
spin-up component from both the ($-p$) $J_{\theta} = +1/2$ component of $\psi_t$ and 
the ($-p$) $J_{\theta} = -1/2$ component of $\psi_t$, and similarly for the $H\omega$ 
operation on the ($+p$) components. It is straightforward to verify that the $A_{dnup\theta}$ 
and $A_{upup\theta}$ calculations produce the same two results. 

On squaring all four of the amplitudes to evaluate the expectation values of the measured 
spins, the results from Eqs.(\ref{eq:ud},\ref{eq:dd})  are each doubled which produces the 
probabilities 
\bqry
P_{(updn+dnup)\omega} & = & \frac{1 + {\rm cos}(\omega)}{2}  \label{eq:prob} \\
P_{(dndn+upup)\omega} & = & \frac{1 - {\rm cos}(\omega)}{2} 
 \eqry
as the required integration over $\theta$ simply reproduces the normalization factor of unity. 

We contend that this demonstrates that the measurement correlations, including those 
discussed by Bell and measured with photons by Clauser and Freedman, Aspect, Zeilinger 
and others~\cite{CF, Aspect, Zeilinger}, depend on the relative orientations of the detectors 
in a way that does not depend on any projection of a state component, but rather on the 
physical quantities that are conserved. 

\section{Linear polarization}

The calculation for detecting coincidence relations between linear polarizations of the photons 
proceeds similarly to the massive neutral particle case. Since the detectors are designed to 
determine the line orientation of linear polarization, it is useful to change from the chiral/helicity 
basis for the entangled photons and switch to a linear polarization basis. Again similarly to the 
massive neutral particle case, this involves describing the entangled photons in the transverse 
plane in terms of the direction of their electric field vectors at the moment of their creation. And 
since at that time there is no net local charge, the total $\vec{E}$-field must vanish, whatever 
charge separation existed prior to the $e^+e^-$ annihilation. It is this total $\vec{E}$-field that is the conserved quantity (due to charge conservation) in this representation. 

Since the total electric field is zero at the origin of the two photons, the orientation of the separate 
$\vec{E}$-field vectors describing the two opposite direction propagating entangled photon fields 
must be $180^o$ out of phase with the other. This is displayed by the equations 
\bqry
|\vec{E}_a> & = & E |{\rm cos}(\theta) \hat{x} + {\rm sin}(\theta) \hat{y}> \nonumber \\
|\vec{E}_b> & = & E |{\rm cos}(\theta+\pi) \hat{x} + {\rm sin}(\theta+\pi) \hat{y} >  \label{eq:E}
\eqry
where $E$ is the common modulus of the separate fields at $z=0$. This must be multiplied 
by the exponentials that describe the propagation of the linearly polarized photons (a,b) in 
time and opposite $z$-directions in space. The exponentials are not explicitly shown as they 
are only relevant to the amplitude at the detector and not the transverse spatial direction of 
significance to the linear polarization filter before the photon detector. (The instantaneous 
amplitude is irrelevant as the detector has a spatial and temporal extent larger in both the 
wavelength and (inverse) frequency of the light being detected) 

The analysis proceeds as in the massive neutral particle case but is slightly complicated by 
 the absence of a dependence on whether the electric field vector is positively or negatively 
 aligned with either direction defined by the angular orientation of the polarizer. Two cases 
 need to be examined, but they yield the same result. The entangled photons state is 
\bqn
|\psi_{\ell}> = \frac{1}{\sqrt{4\pi}} \int d\theta |\vec{E}_a, -p>|\vec{E}_b,+p> 
                              + |\vec{E}_a, +p>|\vec{E}_b,-p>
 \eqn
 and the linear polarizers pass 
 \bqn
<\phi_D(+\omega)| = <\hat{x}, -p |<{\rm cos}(\omega) \hat{x} + {\rm sin}(\omega) \hat{y}, +p| 
\eqn
to the detectors and also, since the polarizers are not sensitive to the spatial direction of the 
electric field
\bqn
<\phi_D(-\omega)| = <-\hat{x}, -p |<-{\rm cos}(\omega) \hat{x} - {\rm sin}(\omega) \hat{y}, +p| 
 \eqn
 where $\omega$ is taken as a positive quantity and the sign of $p$ indicates which is the 
 direction of the photon that is being detected, {\it i.e.}, the location of the polarized filters 
 and detectors relative to the source.
 
 The calculation of the amplitudes and probabilities of the linearly polarized photon 
 detections proceeds similarly to the case of the neutral particles with spin, but now 
 because the directionality (\LQ up" or \LQ down") of the electric field vector is immaterial, 
 both terms produce the same contribution and the probability is as in Eq.(\ref{eq:prob}), 
 {\it i.e.}, the conventional result.
 
 \section{Discussion}

\subsection{Body-fixed frame}

In the Born-Oppenheimer~\cite{BO} (BO) body-fixed frame method, the dipolar charge 
distribution in a water molecule as discussed by Pauling~\cite{LP} can be readily computed. 
However, the electric dipole of the quantum state of a water molecule must vanish due to the 
absence of CP-violation in Quantum Electro-Dynamics. This is accomplished by constructing 
a state that is summed over all orientations of the Born-Oppenheimer body-fixed frame. The 
dipole is only apparent when an electric field is applied which breaks the symmetry and provides 
a preferred frame. Similar considerations can and have been successfully applied to nuclear 
physics~\cite{FV}, including by making use of the quark substructure of nucleons~\cite{BGS} 
parallel to the nucleus and electron structure of atoms. 

This is precisely the character of the spinors of the state of the two entangled neutral particles
discussed here if they were formed in a magnetic field with direction at the angle $\theta$, still 
subject to $J_{\theta}^{tot} = 0$. If one thinks of that as an analog to a body-fixed frame, the total 
amplitude in the absence of the magnetic field is the sum over the full set of \LQ body-fixed frames" 
allowed or required. The same view is taken here of the electric field vectors for linearly polarized 
photons. 

\subsection{Application to EPR}

Hence, an accurate wave function for a more than single particle state may be achieved by 
averaging over the appropriate orientations of body-fixed frame wave functions with an internal 
structure. Applying this, we see that the above total spin zero entangled state of two particles 
with spin may be constructed from wave functions in which the two particles each have their 
spins or electric field vectors (oppositely) aligned along the same particular axis of polarization, 
whether taken parallel or perpendicular to the back-to-back momentum vectors of the pair, and 
then, where necessary, averaging over all orientations of that axis (summing and normalizing). 
Before the projection of the initial $J_{tot}^2 = 0$ by drawing the two particles off along the line 
of a particular direction, averaging over the particular momentum directions would also have been 
required to preserve the initial rotational invariance. 

From this point of view, we have displayed a local quantum reality (which does not include any 
hidden variables) that is manifest and the detection of the spin orientation of either particle simply 
amounts to identifying, specifying and separating out for examination, i.e., preparing, the state 
that is being observed by having projected out only a particular part of the initial total wave function 
for examination and measurement. That part includes the entangled, directly oppositely polarized 
other particle due to either spin conservation with $J_{\theta} = 0$ or the absence of net total 
charge, {\it i.e.}, $\vec{E}_{tot}=0$, for each value of $\theta$. A projection on the appropriate plane 
is accomplished by the selection of pairs of particles that depart the initial state along specific 
spatial directions defined by the back-to-back momenta of the particles.

What needs to be recognized, however, is that the state of the photons that has actually been 
prepared is {\em not} a state with ${J^2_{tot}}= 0$ but rather one with \LQ body-fixed frame" 
components with $J_{z} = 0$ where $z$ is the axis of the photon back-to-back motion or with 
$E_{tot} = 0$ ($J^{tot}_{\theta} = 0$ for the neutral spin-1/2 particles with magnetic moments) 
along the axis defined by $\theta$ and summed over all values of $\theta$. This has transpired 
because a direction of emission for the pair of entangled particles has been determined -- the 
experiment does not accept the entire state into the measurement apparatus, but only the 
particle from one (`body-fixed frame') component in a specific direction and so, due to momentum 
conservation, the apparatus must also only accept the other particle that is propagating in the 
directly opposite direction with the exact opposite transverse electric field or spin direction.

\subsection{Application to Bell's Inequalities}

The significant complication is that the measurement of the orientation of either the spin or 
electric field vector of the two particles does not necessarily mean that the measurement axis 
is precisely aligned with either the common spin direction or electric field vector of the pair. 
This is where Bell's Inequalities develop; those inequalities (contrary to Bell's own expressed 
preference) produce correlations regarding the spin or field directions that are consistent with 
quantum mechanics but exceed those possible in the classical realm. We have displayed two 
special cases where the alignment is precisely known that demonstrate the point straightforwardly, 
but the vanishing of the $\theta$ dependence and the presence of the $\omega$ dependence 
confirms that the correlation, while determined by the entanglement, has nothing to do with 
information being transmitted from one measurement to the other. It depends on the existence 
of a conserved quantity in the initial formation of the entangled state. These explicit calculations 
make clear what is more difficult to discern from Bell's more abstract mathematics but, of course, 
they can only suggest the general result by inference and agreement with Bell's more general 
analysis.

\subsection{Application to Quantum Cryptography}

What conservation law applies in quantum cryptography? This is a conservation law 
constructed by the Hamiltonian of the apparatus that produces the entangled state. 
The two-particle entangled state produced has the structure 
\bqn
\Psi = \frac{1}{\sqrt{2}} [ \; |0>|0> \pm |1>|1> \; ]
\eqn
which has the same character as the para-positronium two photon final state described 
above, namely both particles must have the same quantum number (with the relative sign 
dependent on whether the particles are bosons or fermions). The Hamiltonian constructs 
entangled states with the same character as the `check bit' in a classical computer.

\section{Conclusion}

As the scattering of two particles does produce entanglement, there is a relation between the 
two particles in all parts of the entangled wave function. However, the component of the state 
that will be observed has not been fully prepared. By measuring the position or momentum of 
one of the two, a projection is made onto that particular basis so the prepared state is not the 
initial one about which less is known. Furthermore, there is nothing to prevent measuring the 
position of one particle and the momentum of the other. As shown above, however, this is no 
different in principle than preparing the state with one particular spin orientation using one particle 
but measuring the spin of the other with respect to a different axis. If the precise momentum of 
one is known, so is that of the other, but measuring its location is just equivalent to projecting 
that prepared state wave function onto a different axis. The same is true if the precise relation 
between the spins or electric field vectors of the two particles is known. 

The above shows complete agreement with N.\  David Mermin's description of the effect 
of entanglement~\cite{NDM} as "the closest thing we have to magic". Indeed, one might even 
go so far as to say it is typical of the magician's basic trick, namely, directing attention away 
from what is actually transpiring. 

\section{ Acknowledgments}

I thank Walter Isaacson for his lucid review, in his biography of Einstein, of the history which 
provided the opportunity for these observations to occur. I thank Gerard J.\ Stephenson, Jr.\ 
for discussions and referring me to the para-positronium example.

\end{document}